\documentstyle[aps,prb,eqsecnum,epsfig]{revtex}

\begin{document}

\wideabs{
\title{Surface scaling behavior of isotropic Heisenberg systems: Critical
exponents, structure factor, and profiles}
\author{Michael Krech}
\address{Fachbereich Physik, Bergische Universit\"at - GH
 Wuppertal, 42097 Wuppertal \\ Federal Republic of Germany}
\maketitle

\begin{abstract}
The surface scaling behavior of classical isotropic Heisenberg magnets
is investigated by Monte - Carlo methods in $d=3$ dimensions for various
values of the surface - to - bulk coupling ratio $J_1/J$. For $J_1/J
\leq 1.0$ critical behavior according to the ordinary surface universality
class is found. New estimates for magnetic surface exponents are presented
and compared to older estimates and their theoretical counterparts. For
$J_1/J \geq 2.0$ scaling is still valid with effective exponents which
depend on $J_1/J$. The surface structure factor $S_1(p,L)$ is investigated
at bulk criticality as function of the momentum transfer $p$ parallel to the
surface and the system size $L$. For $J_1/J \leq 1.0$ and $J_1/J \geq 2.0$
the full $p$ dependence of $S_1(p,L)$ can be captured by generalized shape
functions to a remarkable accuracy. Profiles of the magnetization and the
energy density also confirm scaling, where for $J_1/J \leq 1.0$ the ordinary
surface universality class is recovered and for $J_1/J \geq 2.0$ scaling
with $J_1/J$ dependent exponents is found. For $J_1/J = 1.5$ the system
displays a striking crossover behavior from spurious long - range surface
order to the ordinary surface universality class. For $J_1/J \geq 2.0$ the
effective scaling laws must be interpreted as nonasymptotic and the value
$J_1/J = 1.5$ marks a crossover regime, in which the crossover from the
nonasymptotic to the asymptotic (ordinary) surface scaling behavior can be
resolved within numerically attainable system sizes.
\end{abstract}

\draft
\pacs{PACS numbers: 75.10.Hk, 68.35.Rh, 64.60.Fr, 05.70.Jk}
}

\section{Introduction}
Classical spin systems with $O(N)$ symmetry display long - range magnetic
order at low temperatures, if the spatial dimensionality is sufficiently
high or the interaction range is sufficiently large. Long - range
order sets in spontaneously below a certain critical temperature $T = T_c$
at zero magnetic field ${\bf H} = {\bf 0}$, where the field ${\bf H}$
has $N$ components. In the vicinity of the critical point $(T,{\bf H}) =
(T_c,{\bf 0})$ these systems are characterized by an $N$ component order
parameter given by the magnetization. Both from the theoretical and the
experimental point of view these systems provide important realizations
of the well known $O(N)$ universality class for critical behavior. In
particular, the cases $N = 1$, $N = 2$, and $N = 3$ are most relevant for
experiments, because they correspond to the Ising universality class
(magnets with uniaxial anisotropies, simple and binary fluids), the XY
universality class (magnets with planar anisotropies, $^4$He near the
normal - superfluid transition), and the Heisenberg universality class
(isotropic magnets, e.g. Ni, EuO, EuS), respectively. Real samples of any
material always have surfaces which display a distinct critical behavior
in the vicinity of the critical point of the bulk material. In turn, the
penetration depth of surface effects is set by the bulk correlation length
$\xi = \xi^0_\pm |\tau|^{-\nu}$ which {\em diverges} at the critical point 
$\tau = (T-T_c)/T_c \to 0$, where ${\bf H} = {\bf 0}$ is assumed. The exponent
$\nu$ is the {\em universal} correlation length exponent and $\xi^0_\pm$
denotes the nonuniversal correlation length amplitudes above $(+)$ and below
$(-)$ $T_c$ ($N=1$ only). It is now well established that surface
critical behavior obeys the priciple of universality, i.e., critical
exponents and scaling functions which characterize surface critical
behavior do not depend on microscopic details of the material
\cite{Binder83,Diehl86,Diehl97}. General field - theoretic
considerations based on the $\phi^4$ Ginzburg - Landau model for
semiinfinite systems \cite{Diehl86,Diehl97} show that for $O(N)$
symmetric systems there are three surface universality classes which
are denoted as {\em ordinary} $(O)$, {\em surface - bulk} or {\em
special} $(SB)$, and {\em extraordinary} $(E)$. These three
universality classes can be briefly characterized by the absence or
presence of surface order at the critical point of the bulk
material. In particular, $O$ stands for a {\em disordered} surface and
$E$ symbolizes an ordered surface, whereas $SB$ indicates that surface
and bulk are critical simultaneously. The physical nature of the $SB$
surface universality class is that of a {\em multicritical point}
\cite{Diehl86}. Surface order can arise spontaneously $(E)$ or can be
imposed from outside by a surface field ${\bf H}_1$. The latter case
is quite common in fluid systems with confining walls (Ising
universality class, $N = 1$) \cite{FisUp90}, where the bulk transition
is then denoted as {\em normal transition}. It has been shown by rigorous
arguments, that apart from corrections to scaling the normal and the
extraordinary transition are equivalent \cite{BurkDie94}. This is also
quite important for the theory, because the normal transition gives access
to the extraordinary surface universality class even when the surface does
not order spontaneously, i.e., the $SB$ multicritical point does not exist.

Within the framework of $O(N)$ symmetric spin models on lattices we will 
only consider the case of short - range interactions which is characterized
by a nearest - neighbor coupling constant $J$ in the bulk and by another
nearest - neighbor coupling constant $J_1$ in the surface. In this case the
absence or presence of the $SB$ transition crucially depends on the spatial
dimension $d$ and the value of $N$. In $d = 2$ dimensions the $SB$ transition
does not exist, because the "surfaces" in this case are only one - dimensional.
Note that for $N \geq 2$ a $O(N)$ spin system does not show long - range order 
in $d = 2$ \cite{MermWag66}. In $d = 3$ the value of $N$ becomes crucial. For
$N = 1$ the $SB$ transition exists, because the surface, i.e., the Ising model 
in $d = 2$, exhibits long - range order below a finite critical temperature.
In the phase diagram of a semi - infinite Ising ferromagnet the $SB$
multicritical point is located at $T = T_c$, ${\bf H} = {\bf H}_1 = 0$
($N = 1$ here), and the critical value $J_1^c / J$ of the surface - to - bulk
coupling ratio $J_1 / J$. Note that besides the lines of the $O$ and $E$
transitions a line of {\em surface transitions} terminates in the $SB$
multicritical point \cite{Binder83,Diehl86}. For the Ising ferromagnet one
finds $J_1^c / J = 3/2$ on a simple cubic lattice to a remarkably high
accuracy \cite{RuWa95}. For $N = 2$ the surface of the spin system is a
two - dimensional XY model which does no longer show long - range order
\cite{MermWag66}. Instead, a line of Kosterlitz - Thouless surface transitions 
occurs in the $(T, J_1/J)$ plane of the phase diagram \cite{PecLan91} which
terminates in a $SB$ multicritical point at $T = T_c$ and $J_1/J = J_1^c/J
\simeq 1.5$ for a XY ferromagnet on a simple cubic lattice. For $N = 3$ the
$SB$ transition does not exist, because the two - dimensional Heisenberg model 
has no phase transition \cite{MermWag66}. In the absence of symmetry
breaking surface interactions the isotropic Heisenberg ferromagnet in
$d = 3$ should therefore always display {\em ordinary} surface
critical behavior, regardless of the value of $J_1 / J$. In $d \geq 4$
dimensions, i.e., within mean - field theory, the $SB$ transition
exists for any value of $N$.

The dimensional crossover behavior of the $SB$ multicritical point outlined
above greatly complicates the field - theoretic description of the $SB$
transition for general $N$, apart from additional Goldstone problems
below $T_c$ for $N > 1$ \cite{MYu95} which can be accounted for in finite
systems with periodic boundary conditions \cite{CDS96}. Therefore, most
of the field - theoretic work is devoted to the Ising universality class
and to the ordinary transition for general $N \geq 1$
\cite{Diehl86,DieDie81a,DieSh9498}. In the following we will focus on
the ordinary transition in isotropic Heisenberg magnets and we
therefore only summarize some aspects of the ordinary surface
universality class for later reference. For a recent general survey of
surface critical behavior the reader is referred to Ref.\cite{Diehl97}.

Apart from the standard bulk critical exponents additional surface critical
exponents must be introduced in order to describe critical behavior at the
ordinary transition. Let $m_1$ denote the modulus of the surface magnetization
per spin ${\bf M}_1 / \mbox{Area}$ and let $H$ and $H_1$ denote the moduli of
the bulk and surface field $\bf H$ and ${\bf H}_1$, respectively. Then the
critical behavior of $m_1$ and of the susceptibilities $\chi_1$ and
$\chi_{11}$ is given by \cite{Binder83,Diehl86,DieDie81a}
\begin{eqnarray}
\label{mchichi}
m_1 & \sim & (-\tau)^{\beta_1} \nonumber \\
\chi_1 = {\partial m_1 \over \partial H} & \sim & |\tau|^{-\gamma_1} \\
\chi_{11} = {\partial m_1 \over \partial H_1} &\sim & |\tau|^{-\gamma_{11}}
\nonumber
\end{eqnarray}
for $H = H_1 = 0$ and sufficiently small $\tau$. The suceptibilities $\chi_1$
and $\chi_{11}$ are known as the layer susceptibility and the surface or
local susceptibility. In terms of the coordinates ${\bf x}_{||}$ parallel
to the surface and $z$ perpendicular to the surface the correlation decay
exponents $\eta_{||}$ and $\eta_\perp$ are defined by \cite{Binder83,Diehl86}
\begin{eqnarray}
\label{etas}
G(|{\bf x}_{||} - {\bf x'}_{||}|,z,z) & \sim &
|{\bf x}_{||} - {\bf x'}_{||}|^{-(d-2+\eta_{||})} \\
G(0,z,z') & \sim & |z - z'|^{-(d-2+\eta_\perp)} \nonumber
\end{eqnarray}
in the limit of large distances, where $T = T_c$ and $H = H_1 = 0$ is assumed.
The surface exponents defined in Eqs.(\ref{mchichi}) and (\ref{etas}) are
not independent. By virtue of the scaling relations \cite{Binder83,Diehl86}
\begin{eqnarray}
\label{scalrel}
\beta_1 &=& \nu (d - 2 + \eta_{||}) / 2 \nonumber \\
\gamma_1 &=& \nu (2 - \eta_\perp) \\
\gamma_{11} &=& \nu (1 - \eta_{||}) \nonumber \\
\eta_\perp &=& (\eta + \eta_{||}) / 2 \nonumber
\end{eqnarray}
only one of the surface exponents defined above, say, $\eta_{||}$ is
independent for given $\nu$ and $\eta$, where $\eta$ is the decay exponent
of the bulk correlation function according to $G(|{\bf x}^{(1)}-{\bf x}^{(2)}|)
\sim |{\bf x}^{(1)}-{\bf x}^{(2)}|^{-(d-2+\eta)}$ at $T = T_c$. The
remaining magnetic exponents $\delta_1$ and $\delta_{11}$ defined by
\begin{eqnarray}
\label{deltas}
m_1(T_c,H,H_1 = 0) & \sim & H^{1/\delta_1} \\
m_1(T_c,H = 0,H_1) & \sim & H_1^{1/\delta_{11}} \nonumber
\end{eqnarray}
are related to the bulk exponents $\nu$ and $\eta$ and the surface exponent
$\eta_{||}$ according to $\delta_1 = \nu (d+2-\eta) / (2\beta_1)$ and
$\delta_{11} = \nu (d-\eta_{||}) / (2\beta_1)$, respectively
\cite{Binder83,Diehl86}. In analogy with Eq.(\ref{mchichi}) the layer
specific heat $C_1$ and the surface or local specific heat $C_{11}$
display the critical singularites \cite{Binder83,Diehl86}
\begin{eqnarray}
\label{alphas}
C_1 = {\partial e_1 \over \partial t} & \sim & |\tau|^{-\alpha_1} \\
C_{11} = {\partial e_1 \over \partial c} & \sim & |\tau|^{-\alpha_{11}},
\nonumber
\end{eqnarray}
where $e_1$ is the surface energy density per spin and $c = (J_1^c - J_1)/J$
is the surface enhancement. Within the framework of the field - theoretic
renormalization group it has been shown by an explicit calculation of
the energy density profile \cite{DieDie81b} and from the short - distance
expansion of the energy density operator near the surface \cite{DDE83,BurCar87}
that $\alpha_1 = \alpha - 1$ and $\alpha_{11} = \alpha - 2 - \nu$ at the
ordinary transition. The critical singularities of all bulk and surface
quantities can therefore be expressed by the bulk exponents $\nu$ and $\eta$
and the surface exponent $\eta_{||}$, i.e., there is only a single
independent surface exponent at the ordinary transition \cite{Diehl86}.
Within the field - theoretic renormalization group in $d = 4 - \varepsilon$
all surface exponents are known to two - loop order for general $N$
\cite{Diehl86,DieDie81a}. The extrapolation of the $\varepsilon$ -
expansion to $d = 3$ $(\varepsilon = 1)$ already yields reasonable
estimates for the surface exponents, however, their numerical accuracy
remains onknown due to the lack of higher order results which are needed
to apply resummation techniques. Therefore, alternative analytic
approaches have been pursued in order to obtain improved estimates. In
particular for the Heisenberg model $\beta_1 = 0.81 \pm 0.04$ has been
obtained \cite{Ohno84} which agrees with the extrapolated value for
$\beta_1$ \cite{Diehl86}. The result of a perturbative calculation for
the nonlinear sigma model in $d = 2+\varepsilon$ has been combined
with the results in $d = 4 - \varepsilon$ by means of a Pad\'e
approximant \cite{DieNu86} resulting in the numerical estimates
$\beta_1 = 0.84 \pm 0.01$ and $\eta_{||} = 1.39 \pm 0.02$. A massive
field-theoretic approach recently provided the new estimates $\beta_1
= 0.880$, 0.862, 0.889 from a [2/0], [0/2], and [1/1] Pad{\'e}
approximant in $d = 3$, respectively \cite{DieSh9498}. From an
earlier experiment on Ni(100) and Ni(110) surfaces \cite{Alvarado82}
the average estimates $\beta_1 \simeq 0.81$ and $\beta_1 \simeq 0.79$
have been obtained, respectively.

For the experimental verification of surface critical behavior
\cite{Alvarado82} the struture factor in surfaces and thin films is of
particular interest \cite{Dosch92}. From Eq.(\ref{etas}) one concludes
that for small parallel momentum transfer $p$ the surface correlation
function displays the algebraic $p^{-1+\eta_{||}}$ singularity, which
gives access to the exponent $\eta_{||}$ in a surface scattering
experiment. Consequently, substantial theoretical effort has been
spent on the theoretical understanding of correlation functions in
surfaces and films \cite{NF86} and the crossover behavior between them
\cite{NWF87}. A thorough survey of the properties of the static
structure factor can be found in Ref.\cite{KD99}. Scattering
experiments also give access to the order parameter profile which is
governed by universal shape functions. In $d = 2$ and at $T = T_c$ the
shape functions of the order parameter and the energy densitiy profile
for Ising and Potts models can be obtained from conformal invariance
considerations \cite{BurXue91}. Generalizations to $O(N)$ symmetric
models in $d = 2$ are also possible \cite{BurkEi94}. In higher
dimensions one has to resort to field - theoretic methods
\cite{DieDie81b,EKD93}, where only the energy density profile is
nonzero at the ordinary transition \cite{DieDie81b,KED95}. The
identification of the surface universality class for a specific system
is a rather delicate problem. For example, the presence of weak
surface fields leads to a {\em crossover} from ordinary to extraordinary
behavior as the surface is approached from the interior
\cite{RiCzer96}. For multi - component systems, e.g., binary alloys
the surface universality classes may even be different for different
{\em orientations} of the surface \cite{Drew97}.

Whereas most of the theoretical results quoted above can be applied to
the case $d = 3$, $N = 3$ under consideration, numerical
investigations have primarily focussed on the Ising universality class
$(N = 1)$ in $d = 3$. From a Monte - Carlo simulation of the isotropic
Heisenberg ferromagnet on a simple cubic lattice with an open surface 
on one side and a self - consistently determined surface field on the
opposite side \cite{MKB72} an estimate $\beta_1 = 0.75 \pm 0.10$
has been found for the surface exponent $\beta_1$ (ordinary
transition) \cite{BiHo74}. The field dependence of the
surface magnetization $m_1$ at $T = T_c$ (see Eq.(\ref{deltas})) was
determined later \cite{HMK76} and the estimate $\delta_1 = 2.3 \pm
0.1$ was found. Transfer matrix Monte - Carlo calculations for the
Heisenberg ferromagnet on small lattices provided the estimate
$\beta_1 = 0.80 \pm 0.03$ \cite{NiBlo88}. Most of the numerical effort
has been spent on the surface critical behavior of the Ising model
\cite{LanBi90} and a few detailed studies also exist for the XY model
\cite{LanPauBi89}. The most accurate numerical estimates of Ising
surface exponents and amplitude ratios at the ordinary and the SB
transition can be found in Ref.\cite{RuWa95}. We close this brief
overview with the remark that the surface exponent $\beta_1$ for the
ordinary transition is not affected by surface bond disorder or the
presence of steps on the surface \cite{PleiSe98}. A theoretical
explanation for this behavior can be found from the construction of
upper and lower bounds on $\beta_1$ \cite{Diehl98}.

Although a substantial wealth of information is already available for
the surface critrical behavior of the isotropic ferromagnet with
$O(N=3)$ symmetry at the surface, the resulting estimates for the
surface critical exponents are too disparate to provide a reliable
basis for further investigations, e.g, the surface contribution to the
dynamic structure factor. Furthermore, a systematic scaling analysis
of static surface correlations is still missing, which provides
essential information for the numerical analysis of experimental
scattering data and Monte - Carlo data of the dynamic structure
factor. Moreover, previous investigations were focussed on the {\em
asymptotic} scaling regime of the ordinary surface universality class.
In real systems or computer simulations of isotropic Heisenberg
magnets, however, the surface properties at hand may be such that
asymptotic scaling is only obtained after a crossover regime of a
a certain width has been traversed. This also means that some sort of
nonasymptotic surface behavior must occur before the crossover regime
is reached even if the bulk is already critical, i.e., displays
asymptotic scaling behavior governed by bulk exponents. It is
important for the analysis of numerical data to localize these regimes
and to describe their properties.

It is the purpose of this investigation to fill at least some of the
aforementioned gaps. In particular new independent estimates for the
surface critical exponents of the isotropic Heisenberg ferromagnet at
the ordinary transition are provided. With Refs.\cite{KD99} and
\cite{KED95} as guidelines for the universal form of the shape
functions the surface structure factor and the order parameter and
energy density profiles are analyzed at $T = T_c$ for various values
of the surface - to - bulk coupling ratio $J_1/J$. Particular
attention will be paid to the crossover regime, which is best
described by the shape crossover of order parameter and energy density
profiles. Outside the crossover regime a nonasymptotic scaling regime
is investigated and characterized by nonuniversal $J_1/J$ dependent
exponents. This work is mainly focussed on scaling properties and
therefore the simulations are restricted to simple cubic lattices with
(100) surfaces.

\section{Model and simulation method}
The model Hamiltonian describes an isotropic Heisenberg ferro - or
antiferromagnet on a simple cubic lattice with two (100) surfaces
in a cubic geometry. It is given by (see also Ref.\cite{RuWa95})
\begin{equation}
\label{Hamil}
{\cal H} = - J \sum_{\langle i,j \rangle \in V \backslash A}
{\bf S}_i \cdot {\bf S}_j - J_1 \sum_{\langle i,j \rangle \in A}
{\bf S}_i \cdot {\bf S}_j,
\end{equation}
where $V$ denotes the set of all lattice sites (volume) and $A$ denotes the
set of surface sites. The coupling constants $J$ and $J_1$ are assumed to
have the same sign and the spins ${\bf S}_i = (S^x_i,S^y_i,S^z_i)$ with
$|{\bf S}_i| = 1$ only interact with nearest neighbors. A nearest neigbor
pair $\langle i,j \rangle$ of spins is part of the interior $V \backslash A$
of the system, if at least one of the spins is not part of the surface $A$.
A nearest neighbor pair $\langle i,j \rangle$ is part of the surface if both
spins belong to $A$. The surfaces are defined by the lattice planes $(x,y,z=1)$
and $(x,y,z=L)$, where $1 \leq x,y \leq L'$. Along the $x$ and $y$ directions
periodic boundary conditions are applied. In order to avoid unwelcomed
finite - size effects the geometry of the lattice is chosen as cubic,
i.e., $L' = L$. The cubic shape turns out to be a reasonable compromise
between achievable system sizes $L$ and the sensitivity of the system to
the surface - to - bulk coupling ratio $J_1/J$.

The Monte - Carlo algorithm is chosen as a hybrid scheme which consists
of Metropolis sweeps, Wolff single cluster updates \cite{Wolff89},
and overrelaxation sweeps \cite{CFL93}. Typically, one hybrid Monte -
Carlo step consists of 10 individual steps each of which can be one of
the updates listed above. The Metropolis and the Wolff algorithm work the
standard way, where the reduced coordination number of the lattice at the
surfaces and the modified surface coupling $J_1$ must be taken into account.
The acceptance probability $p$ of a proposed spin flip in the Metropolis
algorithm is defined by $p(\Delta E) = 1/[\exp(\Delta E / k_B T) + 1]$,
where $k_B$ is the Boltzmann constant and $\Delta E$ is the change in
configurational energy of the proposed move. The overrelaxation part of
the algorithm performs a microcanonical update of the configuration by
sequentially rotating each spin in the lattice such that its energy
contribution to the energy of the whole configuration remains constant.
The implementation of this update scheme is straightforward, because
according to Eq.(\ref{Hamil}) the energy of a spin with respect to its
neighborhood has the functional form of a scalar product. The angle of
rotation can be chosen randomly for each spin, however, it turns out that
in view of minimal autocorrelation times a reflection, i.e., a rotation
of all spins by 180 degrees is the most efficient overrelaxation move.
One hybrid Monte - Carlo step consists of two Metropolis (M) for single
cluster Wolff (C) and four overrelaxation (O) updates. The individual
updates are mixed automatically in the program to generate the update
sequence M\ O\ C\ O\ C\ M\ O\ C\ O\ C\ . The random number generator is
the shift register generator R1279 defined by the recursion relation
$X_n = X_{n-p} \oplus X_{n-q}$ for $(p,q) = (1279,1063)$. Generators like
these are known to cause systematic errors in combination with the Wolff
algorithm \cite{cluerr}. However, for lags $(p,q)$ used here these errors
are far smaller than typical statistical errors. They are further reduced
by the hybrid nature of the algorithm \cite{AFMDPL}.

The Monte - Carlo scheme described above is employed for lattice sizes
$L$ between $L = 12$ and $L = 72$. For each choice of parameters we perform
at least 20 blocks of $10^3$ hybid steps for equilibration followed by
$10^4$ hybrid steps for measurements. Each measurement block yields an
estimate for all quantities of interest and from these we obtain our
final estimates and estimates of their statistical error following
standard procedures. The integrated autocorrelation time of the hybrid
algorithm is determined by the autocorrelation function of the energy
or, equivalently, the modulus of the magnetization, which yield the
slowest modes for the Wolff algorithm. The autocorrelation times do not
exceed 10 hybrid Monte - Carlo steps for the largest lattice size $(T =
T_c)$, so the equilibration and measurement periods quoted above translate
to roughly 100 and 1000 autocorrelation times, respectively. In order to
obtain the best statistics for all magnetic quantities a measurement
is made after every hybrid Monte - Carlo step. All error bars quoted
in the following correspond to one standard deviation. The hybrid
scheme samples the surfaces of the system reasonably often, so a
preferential sampling of surface configurations is not required. The
simulations have been performed on DEC alpha and AXP workstations at the
Physics department of the BUGH Wuppertal.

\section{Surface scaling exponents}
The simulations presented here have been performed at $T = T_c$ for
several values of the surface - to - bulk coupling ratio $J_1/J$. The
estimate for $T_c$ used here is taken from Ref.\cite{CFL93}, where
the critical coupling $K_c$ has been determined as $K_c \equiv J/(k_B T_c)
= 0.693035(37)$. In view of the limited system size $L \leq 72$ the
relative accuracy of $10^{-4}$ in $T_c$ is sufficient in order to
perform a standard finite - size scaling analysis in terms of the
system size $L$ up to the usual corrections to scaling. As our main
reference for bulk critical exponents we choose the work of Guida and
Zinn - Justin \cite{GZJ98}, where the bulk critical exponents of the
$O(N)$ universality class have been obtained from high order Borel
resummed perturbation theory for the Ginzburg - Landau model.

In order to access surface critical exponents the surface magnetization
$m_1$, its second moment, and the surface energy density $e_1$ has been
measured. The layer and surface specific heats $C_1$ and $C_{11}$ (see
Eq.(\ref{alphas})) have not been investgated, because $\alpha_1$ and
$\alpha_{11}$ can be expressed by bulk exponents only. The layer
magnetization ${\bf m}(z)$, $1 \leq z \leq L$ and the total magnetization
${\bf m}_{tot}$ are defined by
\begin{eqnarray}
\label{mzmb}
{\bf m}(z) &\equiv& \sum_{x,y=1}^L {\bf S}_{xyz} / L^2 \nonumber \\ \\
{\bf m}_{tot} &\equiv& \sum_{z=1}^L {\bf m}(z) / L, \nonumber
\end{eqnarray}
respectively, and $m_{tot} \equiv |{\bf m}_{tot}|$ denotes the modulus of the
total magnetization. The magnetization profile $m(z)$ is then defined as
the projection of ${\bf m}(z)$ onto the total magnetization ${\bf m}_{tot}$,
i.e.,
\begin{equation}
\label{mz}
m(z) \equiv {\bf m}(z) \cdot {\bf m}_{tot} / m_{tot}
\end{equation}
and $m_1 \equiv (m(1) + m(L)) / 2$ defines the surface
magnetization. Note that the surfaces at $z=1$ and $z=L$ are identical
(see Eq.(\ref{Hamil})). In terms of $m_{tot}$ and $m_1$ the layer and
surface susceptibilities (i.e., their longitudinal components)
$\chi_1$ and $\chi_{11}$ for a completely finite system are defined as
\begin{eqnarray}
\label{chis}
\chi_1 &=& L^2 (\langle m_{tot} m_1 \rangle - \langle m_{tot} \rangle
\langle m_1 \rangle) / (k_B T) \\
\chi_{11} &=& L^2 (\langle m_1^2 \rangle - \langle m_1 \rangle^2) / (k_B T),
\nonumber
\end{eqnarray}
where $\langle \dots \rangle$ denotes the thermal average. The energy
profile $e(z)$ is defined accordingly, where apart from the exchange energy
between the spins {\em within} layer $z$ half the interaction energy
to the layers $z-1$ and $z+1$ also contributes to $e(z)$, so $e_{tot} \equiv
\sum_{z=1}^L e(z) / L$ is the total energy density. In the following
all energies will be given in units of $k_B T_c$, i.e., extra factors
$k_B T$ (see, e.g, Eq.(\ref{chis})) are unity at $T = T_c$.
\begin{figure}
\centerline{\epsfig{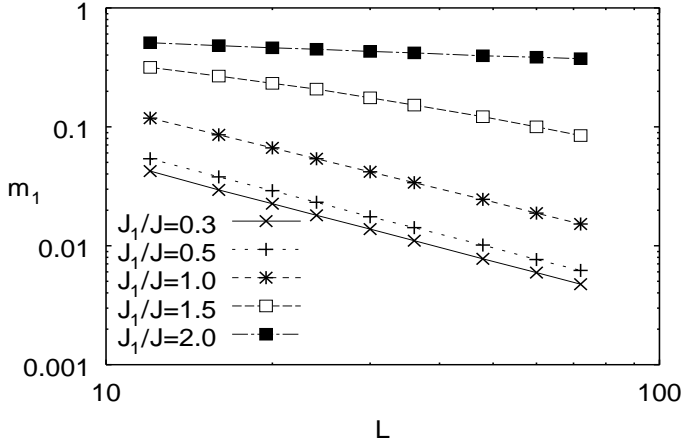}}
\caption{Surface magnetization $m_1(L)$ (see Eq.(\protect\ref{mz})) at
$T=T_c$ as function of the system size $L$ for $J_1/J = 0.3$
($\times$, solid line), 0.5 (+, short dashed line), 1.0 ($*$, dashed
line), 1.5 ($\Box$, long dashed line), and 2.0 (\protect\rule{2mm}{2mm},
dash-dotted line). Statistical errors are much smaller than the symbol
sizes and the various lines are just guides to the eye. For $J_1/J\leq
1$ $m_1$ displays the expected behavior for the ordinary surface
universality class. For $J_1/J=1.5$ the system undergoes a crossover
towards ordinary surface behavior, whereas for $J_1/J=2.0$ the
behavior of $m_1$ is inherently nonasymptotic within the available
range of system sizes.
\label{m1L}}
\end{figure}

\begin{figure}
\centerline{\epsfig{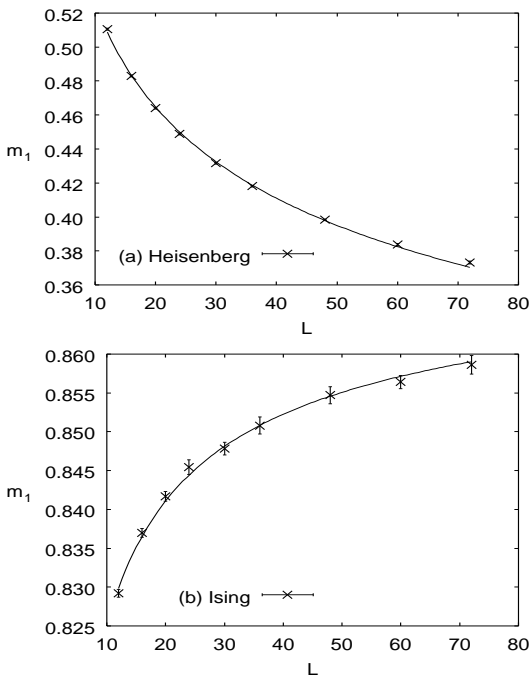}}
\caption{Surface magnetization $m_1(L)$ for $J_1/J = 2.0$ at $T = T_c$
($\times$) (a) for the Heisenberg model given by
Eq.(\protect\ref{Hamil}) (see also Fig.\protect\ref{m1L}) and (b) for
a corresponding Ising model \protect\cite{MKIsing}. For the Heisenberg
model (a) $m_1(L)$ decays according to an effective power law (solid
line), whereas for the Ising model (b) $m_1(L)$ increases towards the
value of the spontaneous surface magnetization according to
Eq.(\protect\ref{m1LI}) (solid line).
\label{m1HI}}
\end{figure}

The surface magnetization $m_1$ at $T = T_c$ as function of the system
size $L$ is shown in Fig.\ref{m1L} for $J_1/J = 0.3$, 0.5, 1.0, 1.5,
and 2.0. For $J_1/J \leq 1$ the functional form of $m_1(L) \equiv
\langle m_1 \rangle$ is accurately captured by
\begin{equation}
\label{m1Lfit}
m_1(L) = B_{m_1} L^{-\beta_1/\nu}\left(1 + C_{m_1} L^{-\omega} \right),
\end{equation}
where $B_{m_1}$ is the magnetization amplitude and $C_{m_1}$ is the amplitude
of the leading correction to scaling. The associated Wegner exponent
$\omega = 0.78$ is taken from Ref.\cite{GZJ98}. A least square fit of
Eq.(\ref{m1Lfit}) to the data for $J_1/J \leq 1$ displayed in
Fig.\ref{m1L} yields the estimates $\beta_1/\nu = 1.185(6)$,
1.175(13), and 1.171(7) for $J_1/J = 0.3$, 0.5, and 1.0,
respectively. The error indicated in parenthesis corresponds to one
standard deviation. From these estimates one obtaines the weighted
average $\beta_1/\nu = 1.179(6)$, where the smallest of the individual
errors is taken as the final error estimate. Two of the three
individual estimates are included in the error interval of the final 
estimate. From the literature value $\nu = 0.7073(35)$ \cite{GZJ98}
one obtains the estimate
\begin{equation}
\label{beta1}
\beta_1 = 0.834(6)
\end{equation}
for the surface exponent of the magnetization. For $J_1/J = 1.5$
Eq.(\ref{m1Lfit}) does not capture the functional form of $m_1(L)$,
because within the available range of lattice sizes the system undergoes
a crossover towards the asymptotic ordinary surface critical behavior.
A more detailed discussion of this crossover is postponed to Sec. 5,
where the order parameter and energy density profiles are presented.
If the decay of $m_1(L)$ for $J_1/J = 1.5$ is described by
an {\em effective} exponent according to Eq.(\ref{m1Lfit}), one
finds a value around 0.6 for $L \leq 20$ and a value around 0.9 for $L
\geq 48$. This indicates that only a part of the full
crossover process is captured by the simulation. This leads to the
conclusion that the data for $J_1/J = 2.0$ have not yet even entered the
crossover regime to ordinary surface critical behavior. It is
insructive to compare these data with corresponding data for the Ising
model. The surface - to - bulk coupling ratio $J_1/J = 2.0$ already
belongs to the extraordinary regime of the Ising model \cite{RuWa95},
where the surface exhibits long - range order at $T = T_c$. The
comparison is shown in Fig.\ref{m1HI}, where the data for an Ising
model according to Eq.(\ref{Hamil}) have been obtained from a hybrid
algortihm which corresponds to the one described above, except that
overrelaxation moves cannot be performed in this case \cite{MKIsing}.
The surface magnetization decays with an effective exponent of about
0.16 (see Fig.\ref{m1HI}(a), solid line), whereas for the Ising model (see
Fig.\ref{m1HI}(b), solid line) $m_1(L)$ approaches the spontaneous surface
magnetization $m_{10}$ according to
\begin{equation}
\label{m1LI}
m_1(L) = m_{10} - B^I_{m_1} L^{-\beta/\nu}
\end{equation}
up to corrections to scaling, where $\beta/\nu \simeq 0.517$
\cite{GZJ98} is the scaling dimension of the order parameter in the
Ising universality class and $B^I_{m_1}$ is a nonuniversal
amplitude. Fig.\ref{m1HI} illustrates how the presence of real long -
range surface order (b) can be distinguished from spurious long - range
surface order (a) which only appears as a nonasymptotic finite - size
effect. However, within a typical range of numerically accessible
system sizes the crossover to the asymptotic ordinary surface behavior
cannot be observed for $J_1/J \geq 2.0$ and therefore a data analysis
within the framework of finite - size scaling only yields an {\em
effective} exponent $\beta_{1,eff} = \beta_{1,eff}(J_1/J)$. For
$J_1/J = 2.0$ one has $\beta_{1,eff}(2.0)/\nu \simeq 0.16$ (see
Fig.\ref{m1HI}(a)) and for $J_1/J = 3.0$ an effective exponent
$\beta_{1,eff}(3.0)/\nu \simeq 0.08$ is found (not shown).
We do not quote error bars here, because the estimates for
$\beta_{1,eff}(J_1/J)$ are presumably affected by systematic errors
which are larger than the statistical ones. In the interior of the
system the 'bulk' magnetization $m_b \equiv \langle m(L/2) \rangle$
obeys standard critical finite size scaling with the exponent
$\beta/\nu \simeq 0.518$ \cite{GZJ98} (see Eqs.(\ref{mscal}) and
(\ref{mb}) and Fig.\ref{mz20} in Sec. 5).

\begin{figure}
\centerline{\epsfig{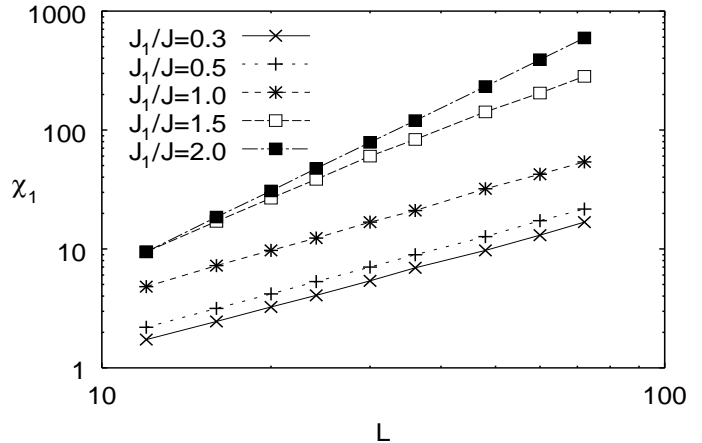}}
\caption{Layer susceptibility $\chi_1$ (see Eq.(\protect\ref{chis})) at
$T=T_c$ as function of the system size $L$ for $J_1/J = 0.3$
($\times$, solid line), 0.5 (+, short dashed line), 1.0 ($*$, dashed
line), 1.5 ($\Box$, long dashed line), and 2.0 (\protect\rule{2mm}{2mm},
dash-dotted line). Statistical errors are smaller than the symbol
sizes and the various lines are just guides to the eye. For $J_1/J\leq
1$ $\chi_1$ displays the expected behavior for the ordinary surface
universality class. For $J_1/J=1.5$ the system undergoes a crossover
towards ordinary surface behavior, whereas for $J_1/J=2.0$ the
behavior of $\chi_1$ is inherently nonasymptotic within the available
range of system sizes (see also Fig.\protect\ref{m1L}).
\label{X1L}}
\end{figure}

The layer susceptibility $\chi_1$ is analyzed in the same way as the
surface magnetization $m_1(L)$. The data are displayed in
Fig.\ref{X1L}. For $J_1/J \leq 1$ the data can be interpreted
according to
\begin{equation}
\label{X1Lfit}
\chi_1(L) = B_{\chi_1} L^{\gamma_1/\nu}\left(1 + C_{\chi_1}
L^{-\omega} \right),
\end{equation}
which is the exact analog of Eq.(\ref{m1Lfit}). The term in
parenthesis captures the leading correction to scaling, $\gamma_1/\nu$
is the corresponding surface exponent for finite - size scaling, and
$B_{\chi_1}$ is a nonuniversal overall amplitude. With $\omega \simeq
0.78$ as above a least square fit of Eq.(\ref{X1Lfit}) to the data
shown in Fig.\ref{X1L} yields the estimates $\gamma_1/\nu =
1.314(23)$, 1.305(12), and 1.308(25) for $J_1/J = 0.3$, 0.5, and 1.0,
respectively. As before we adopt the weighted average $\gamma_1/\nu =
1.307(12)$ as our final estimate, where the smallest of the individual
errors is taken as the error estimate. From the literature value $\nu
= 0.7073(35)$ the estimate
\begin{equation}
\label{gamma1}
\gamma_1 = 0.924(10)
\end{equation}
is found. For $J_1/J \geq 1.5$ the data qualitatively
behave as in Fig.\ref{m1L}: For $J_1/J = 1.5$ the system undergoes a
crossover to ordinary surface behavior and for $J_1/J = 2.0$ the
asymptotic behavior is out of reach for the simulation. Nonetheless,
the increase of $\chi_1$ in the latter case can be described by an
effective exponent $\gamma_{1,eff}(J_1/J=2.0)/\nu \simeq 2.31$. Likewise,
$\gamma_{1,eff}(J_1/J=3.0)/\nu \simeq 2.44$ is obtained (not shown).
Error bars are not quoted for the reasons indicated above.

The exponets $\beta_1$ and $\gamma_1$ are not independent. From the
scaling relations given by Eq.(\ref{scalrel}) and bulk scaling
relations one can infer the simple rule $\beta_1 + \gamma_1 = \beta +
\gamma$. From the literature \cite{GZJ98} one obtains $\beta + \gamma
= 1.7557(56)$ and Eqs.(\ref{beta1}) and (\ref{gamma1}) yield $\beta_1
+ \gamma_1 = 1.758(11)$ which verifies the above scaling law. The
effective exponents $\beta_{1,eff}(2.0)$ and $\gamma_{1,eff}(2.0)$ for
$J_1/J = 2.0$ (see Figs.\ref{m1L} and \ref{X1L}) yield
$\beta_{1,eff}(2.0) + \gamma_{1,eff}(2.0) \simeq 1.75$ which is remarkably
close to the value of $\beta + \gamma$ quoted above. Likewise,
$\beta_{1,eff}(3.0) + \gamma_{1,eff}(3.0) \simeq 1.78$ is found for $J_1/J
= 3.0$. From Eq.(\ref{scalrel}) one furthermore obtains the correlation
exponents
\begin{equation}
\label{etapp}
\eta_{||} = 1.358(12) \quad , \quad \eta_\perp = 0.697(6)
\end{equation}
and the exponent
\begin{equation}
\label{gamma11}
\gamma_{11} = -0.253(9)
\end{equation}
of the surface susceptibility $\chi_{11}$. At $T = T_c$ the surface
susceptibility therefore behaves according to
\begin{equation}
\label{X11Lfit}
\chi_{11}(L) = \chi_{110} - B_{\chi_{11}} L^{1-\eta_{||}} \left(1 +
C_{\chi_{11}} L^{-\omega} \right),
\end{equation}
\begin{figure}
\centerline{\epsfig{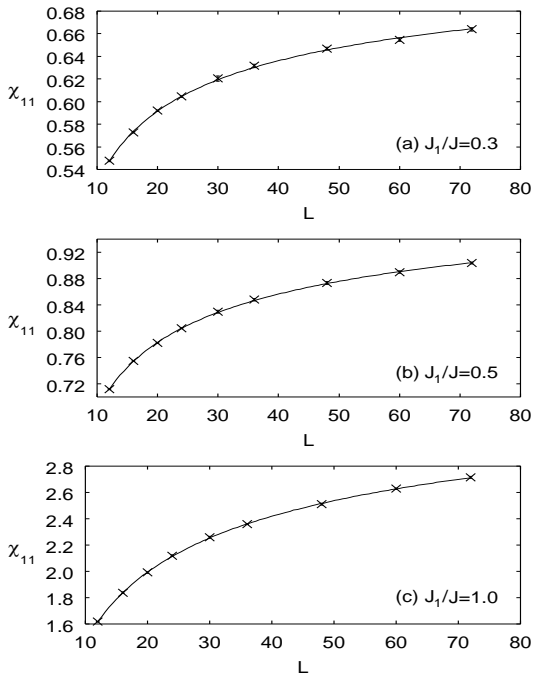}}
\caption{Surface susceptibility $\chi_{11}$ at $T = T_c$ ($\times$) as
function of the system size $L$ for (a) $J_1/J = 0.3$, (b) $J_1/J =
0.5$, and (c) $J_1/J = 1.0$. Error bars are much smaller than the
symbol sizes. The solid lines display fits of
Eq.(\protect\ref{X11Lfit}) to the data (a), (b), and (c), respectively.
\label{X11L}}
\end{figure}
\noindent
\begin{minipage}[t]{8.6cm}
\begin{table}
\caption{Comparison of estimates for $\beta_1$ from experiments on
Ni(100) \protect\cite{Alvarado82} (average of three measurements),
early Monte - Carlo work \protect\cite{BiHo74}, high temperature
series \protect\cite{Ohno84}, field theory \protect\cite{DieNu86},
transfer matrix Monte - Carlo \protect\cite{NiBlo88}, massive field
theory \protect\cite{DieSh9498} ([0/2], [2/0], [1/1] Pad{\'e}
approximants), and Eq.(\protect\ref{beta1}).
\label{beta1tab}}
\begin{tabular}{cccccccc}
\ &\ Ref.\cite{Alvarado82}\ &\ Ref.\cite{BiHo74}\ &\ Ref.\cite{Ohno84}\ &
\ Ref.\cite{DieNu86}\ &\ Ref.\cite{NiBlo88}\ &\ Ref.\cite{DieSh9498}\ &
\ Eq.(\ref{beta1})\ \\
\hline
$\beta_1$ & 0.81 & 0.75(10) & 0.81(4) & 0.84(1) & 0.80(3) &
\begin{tabular}{c} 0.862 \\ 0.889 \end{tabular} & 0.834(6)\\
\end{tabular}
\end{table}
\end{minipage}
where the term in parenthesis captures the leading correction to
scaling. Fits of Eq.(\ref{X11Lfit}) to the data for $J_1/J \leq 1$ are
shown in Fig.\ref{X11L}, where $1-\eta_{||} = -0.358$ and $\omega =
0.78$ are kept fixed and the amplitudes $\chi_{110}$, $B_{\chi_{11}}$
and $C_{\chi_{11}}$ are used as fit parameters. For $J_1/J \geq 1.5$
Eq.(\ref{X11Lfit}) does no longer describe $\chi_{11}$ due to the
strong nonasymptotic effects described above (for more details see
Sec. 4). A comparison of different estimates for $\beta_1$ from
various sources is shown in Table \ref{beta1tab}. The estimates are
rather consistent within their mutual errors, but the best agreement
is found with the field theoretic estimate given in Ref.\cite{DieNu86}.

\begin{figure}
\centerline{\epsfig{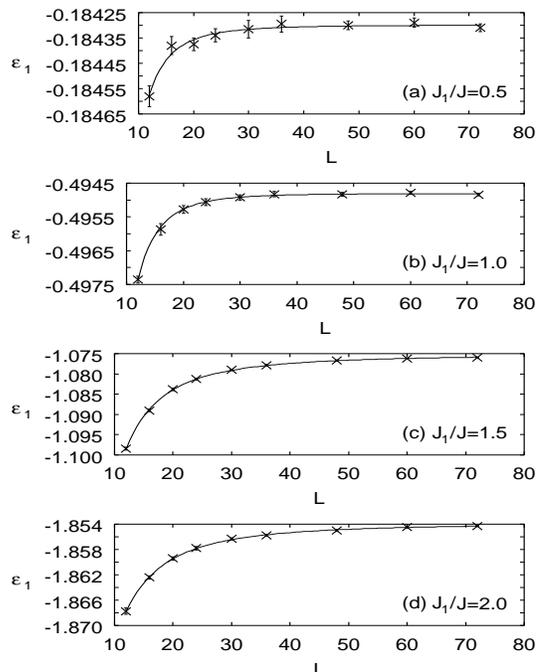}}
\caption{Dimensionless surface energy density $\varepsilon_1 =
e_1/(k_B T_c)$ at $T = T_c$ ($\times$) as function of the system size
$L$ for (a) $J_1/J = 0.5$, (b) $J_1/J = 1.0$, (c) $J_1/J = 1.5$, and
(d) $J_1/J = 2.0$. In (c) and (d) the error bars are much smaller than
the symbol sizes. The solid lines display fits of
Eq.(\protect\ref{e1Lfit}) to the data (a), (b), (c), and (d),
respectively.
\label{e1L}}
\end{figure}

We close this section with an investigation of the finite-size
behavior of the surface energy density $e_1$. The data for the
dimensionless surface energy density $\varepsilon_1 = e_1/(k_B T_c)$
are shown in Fig.\ref{e1L} for $0.5 \leq J_1/J \leq 2.0$. The $L$
dependence of $e_1$ for $J_1/J = 0.3$ (not shown), 0.5 (a), and 1.0
(b) is qualitatively different from the $L$ dependence for $J_1/J =
1.5$ (c), 2.0 (d), and 3.0 (not shown). The leading $L$ dependence of
$e_1$ is written as
\begin{equation}
\label{e1Lfit}
\varepsilon_1(L) = \varepsilon_{10} + B^{(2)}_{e_1} L^{-2} +
B^{(3)}_{e_1} L^{-3},
\end{equation}
where the amplitudes $\varepsilon_{10}$, $B^{(2)}_{e_1}$, and
$B^{(3)}_{e_1}$ are taken as fit parameters. For $J_1/J \leq 1$ the
coefficient $B^{(2)}_{e_1}$ turns out to be two orders of magnitude
smaller than the coefficient $B^{(3)}_{e_1}$ which suggests that
$L^{-3}$ is the leading finite - size correction in this
case. According to naive finite - size scaling one expects $x_{e_1} =
(1 - \alpha_1)/\nu$ as the leading finite - size exponent for the
surface energy density. In fact, due to the scaling relation $\alpha_1
= \alpha - 1$ for the ordinary transition \cite{DDE83,BurCar87} one
finds $x_{e_1} = (2 - \alpha)/\nu = d = 3$ from hyperscaling in
agreement with the simulation data in Figs.\ref{e1L} (a) and (b).
In contrast, $B^{(2)}_{e_1}$ and $B^{(3)}_{e_1}$ are of roughly equal
magnitude (and of opposite sign) for the case $J_1/J \geq 1.5$ displayed in
Figs.\ref{e1L} (c) and (d). Therefore, $L^{-2}$ rather than $L^{-3}$ is
the leading finite - size correction here. The discrepancy between
Figs.\ref{e1L} (a), (b) and Figs.\ref{e1L} (c), (d) is due to strong
nonasymptotic contributions to the finite size behavior for $J_1/J \geq
1.5$. One possible source of the $L^{-2}$ contribution is the scaling
dimension $x_{f_s} = (2-\alpha_s)/\nu = (2-\alpha-\nu)/\nu = d-1 = 2$
of the surface free energy $f_s$ (surface tension) which prevails in
Eq.(\ref{e1Lfit}) due to the {\em nonscaling} dependence of the surface
tension on $J_1/J$ \cite{Privman90}. Another source of the leading $L^{-2}$
dependence is the {\em regular} (noncritical) finite - size behavior of
the surface tension for periodic boundary conditions \cite{Privman90}.
Other finite - size corrections such as $L^{-1}$ substantially reduce the
goodness - of - fit and can therefore be ruled out as leading terms.
Possible logarithmic corrections of the form $L^{-2} \log L$
\cite{Privman90} cannot be identified unambigiously.
\begin{figure}
\centerline{\epsfig{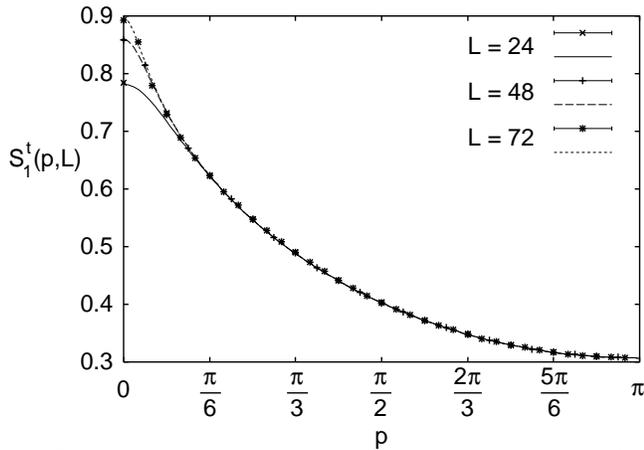}}
\caption{Transverse component of the surface structure factor $S^t_1(p,L)$
for $J_1/J = 0.5$ and $L = 24$ ($\times$), $L = 48$ ($+$), and $L = 72$
($*$) as function of the momentum transfer $p$ in the (100) direction.
Numerical evaluations of Eq.(\protect\ref{S1pfit}) in the limit $c \to
\infty$ for $L = 24$, $L = 48$, and $L = 72$ (see main text) are displayed
by the solid, the long dashed and short dashed line, respectively.
\label{S1p05}}
\end{figure}

\section{Surface structure factor}
The surface structure factor is given by the discrete Fourier transform
of the surface spin - spin correlation function
\begin{eqnarray}
\label{Gx1}
G_1^{\alpha \alpha}({\bf x} - {\bf x}') &=& \left(
\langle S_{x,y,1}^{\alpha} S_{x',y',1}^{\alpha} \rangle
- \langle S_{x,y,1}^{\alpha} \rangle^2 \right. \\
&+& \left. \langle S_{x,y,L}^{\alpha} S_{x',y',L}^{\alpha} \rangle
- \langle S_{x,y,L}^{\alpha} \rangle^2 \right)/2, \nonumber
\end{eqnarray}
where the Fourier transform is taken with respect to ${\bf x} - {\bf x}'$,
${\bf x} = (x,y)$, and the upper index $\alpha$ refers to the spin component.
Note that off - diagonal components of $G_1$ vanish identically. The
longitudinal and transverse components of $G_1^{\alpha \alpha}$ with
respect to the total magnetization are very similar and we therefore
restrict the following discussion to the transverse component only.
The momentum transfer ${\bf p}$ in the (100) direction, i.e., parallel to
the surface, in units of the inverse lattice constant is given by ${\bf p}
= (p,0,0) = (2\pi n/L,0,0)$, $n = 0,1,\dots ,L/2$. Numerical data of
the surface structure factor for $J_1/J = 0.5$ in the (100) direction
for three different lattice sizes are displayed in
Fig.\ref{S1p05}. The data essentially collapse onto a single curve for
$p > 0$, finite - size effects are only visible at $p = 0$, where the
structure factor shown in Fig.\ref{S1p05} reduces to the transverse
component of the surface susceptibility $\chi_{11}$ (see
Eq.(\ref{X11Lfit})). For $J_1/J = 0.3$ identical properties are
obtained (not shown). It turns out that lattice effects in the
structure factor near the Brillouin zone boundary can be captured by
the replacement $p \to 2 \sin(p/2)$ to a remarkable accuracy
\cite{RZW94}. The shape of the surface structure factor is reasonably
well captured by the mean - field type expression \cite{MKIsing,RZW94}
\begin{eqnarray}
\label{S1pfit}
S^{\alpha}_1(p,L) &=& \chi^{\alpha}_{110} - B^{\alpha}_{\chi_{11}}
{c^{\alpha} - 2\sin(p/2) \over c^{\alpha} + 2\sin(p/2)} \\
&\times& \left[ {2 \Gamma^{\alpha} \sin(p/2) \over \tanh \left(2
\Gamma^{\alpha} \sin(p/2) L \right)} \right]^{\eta_{||}-1}, \nonumber
\end{eqnarray}
where the upper index $\alpha$ indicates the component (longitudinal $l$
or transverse $t$ with respect to ${\bf m}_{tot}$) of $S^{\alpha}_1(p,L)$.
The coefficients $\chi^{\alpha}_{110}$ and $B^{\alpha}_{\chi_{11}}$ are
the coefficients of the surface susceptibility $\chi_{11}^{\alpha}$
(for $\alpha = l$ see Eq.(\ref{X11Lfit})).
The width parameter $\Gamma^{\alpha}$ and the surface enhancement
parameter $c^{\alpha}$ are used to fit the momentum dependence of the
surface structure factor and the exponent $\eta_{||}$ is taken from
Eq.(\ref{etapp}). The fit is performed in two stages. First,
$\chi^{\alpha}_{110}$ and $B^{\alpha}_{\chi_{11}}$ are determined from
a least square fit of Eq.(\ref{S1pfit}) for $p = 0$ to the surface
susceptibility. Second, the width parameter $\Gamma^{\alpha}$ and, if
needed (see below), the surface enhancement parameter $c^{\alpha}$ are
adjusted to obtain a least square fit of the momentum dependence to the
data. In practice, this fit procedure has only been performed for the
largest lattice size $L = 72$. For smaller systems the fit parameters
are taken from $L = 72$ in order to test the accuracy of the shape
predicted by Eq.(\ref{S1pfit}). The result for $J_1/J = 0.5$ is displayed in
Fig.\ref{S1p05} for the transverse component $S^t_1(p,L)$ of the
surface structure factor. For the largest system $L = 72 (*)$ the short
dashed line shows the fit, whereas for other lattice sizes such as $L
= 24 (\times)$ and $48 (+)$ the evaluation of Eq.(\ref{S1pfit}) for $L
= 24$ (solid line) and $L = 48$ (long dashed line) confirms the
predicted shape to a remarkable accuracy. For $J_1/J = 0.3$ and
0.5 the parameter $c^{\alpha}$ is very large and therefore the surface
enhancement prefactor $(c^{\alpha} - 2\sin(p/2))/(c^{\alpha} +
2\sin(p/2))$ can be omitted. Noticeable deviations between
the data and Eq.(\ref{S1pfit}) for $\alpha = t$ only occur for small
$p$ on the smaller lattices, where additional lattice corrections to
Eq.(\ref{S1pfit}) may become important. However, for any $p > 0$ the
data for all system sizes collapse onto a single curve, which can be
obtained from Eq.(\ref{S1pfit}) by performing the limits $c^{\alpha}
\to \infty$ and $L \to \infty$ at finite $p$.

\begin{figure}
\centerline{\epsfig{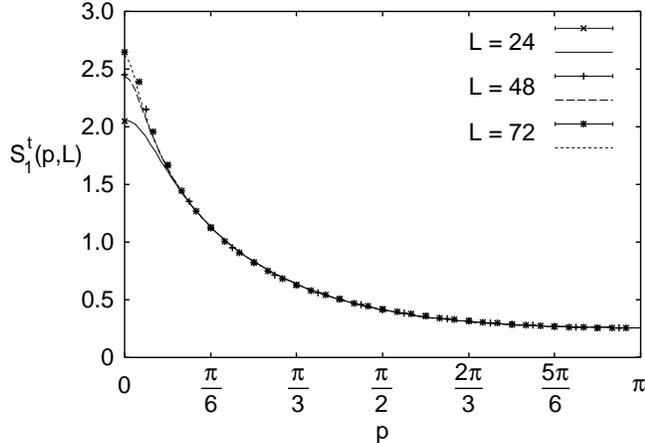}}
\caption{Transverse component of the surface structure factor $S^t_1(p,L)$
for $J_1/J = 1.0$ and $L = 24$ ($\times$), $L = 48$ ($+$), and $L = 72$
($*$) as function of the momentum transfer $p$ in the (100) direction.
Numerical evaluations of Eq.(\protect\ref{S1pfit}) for $L = 24$, $L = 48$,
and $L = 72$ (see main text) are displayed by the solid, the long
dashed and short dashed line, respectively. Surface enhancement
corrections are essential for the data analysis.
\label{S1p10}}
\end{figure}

For $J_1/J = 1.0$ the surface enhancement parameter $c^{\alpha}$
becomes important. The data are shown in Fig.\ref{S1p10}, where the
surface enhancement prefactor provides an important correction to the
momentum dependence of $S^{\alpha}_1(p,L)$. The fit procedure works as
described above, but deviations from the assumed shape for small $p$
also occur for $L = 72$. The surface structure factor is more
sensitive to crossover phenomena occurring for larger values of
$J_1/J$ than the surface quantities discussed in the previous
section. Corrections to scaling, which are not included in 
Eq.(\ref{S1pfit}), may also account for part of the deviations between the
data and the simple model for the shape function. Note that the
analytic results obtained in Ref.\cite{KD99} do not hold for the cubic
geometry used here. Furthermore, nonasymptotic surface enhancement
corrections, which have not been considered in Ref.\cite{KD99}, become
essential for the data analysis. For $p > 0$ the data still collapse
onto a single curve, which is given by Eq.(\ref{S1pfit}) in
the limit $L \to \infty$ at finite $p$ {\em and finite} $c^{\alpha}
\simeq 15$.

\begin{figure}
\centerline{\epsfig{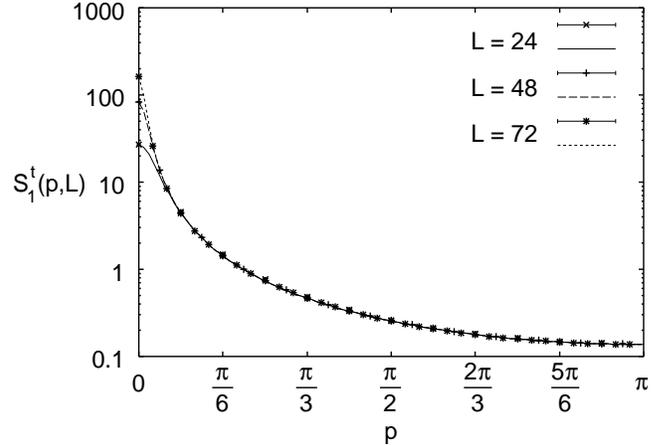}}
\caption{Transverse component of the surface structure factor $S^t_1(p,L)$
for $J_1/J = 2.0$ and $L = 24$ ($\times$), $L = 48$ ($+$), and $L = 72$
($*$) as function of the momentum transfer $p$ in the (100) direction.
Numerical evaluations of Eq.(\protect\ref{S1p20fit}) for $L = 24$, $L = 48$,
and $L = 72$ (see main text) are displayed by the solid, the long
dashed and short dashed line, respectively. The behavior of $S^t_1$ is
is inherently nonasymptotic.
\label{S1p20}}
\end{figure}

The behavior of $S^t_1(p,L)$ for $J_1/J = 2.0$ is shown in Fig.\ref{S1p20},
for $J_1/J = 3.0$ similar results have been obtained (not shown).
Although scaling appears to be valid to a very high degree of accuracy,
the behavior of $S^t_1(p,L)$ is very different from Figs.\ref{S1p05} and
\ref{S1p10}. First, the surface susceptibility {\em grows} according to
the power law
\begin{equation}
\label{X1120fit}
\chi^{\alpha}_{11}(L) = B^{\alpha}_{\chi_{11}} L^{1-\eta_{||,eff}(J_1/J)}
\end{equation}
rather than approaching a finite limit as in Eq.(\ref{X11Lfit}). From a
least square fit of Eq.(\ref{X1120fit}) to $\chi^t_{11}$ one obtains
the effective exponent $\eta_{||,eff}(J_1/J=2.0) \simeq -0.64$, where
deviations from the pure power law given by Eq.(\ref{X1120fit}) are very
small. For $J_1/J = 3.0$ one obtains $\eta_{||,eff}(3.0) \simeq -0.82$.
We refrain from quoting error bars here, because the values for
$\eta_{||,eff}(J_1/J)$ may be affected by systematic errors due to
corrections to Eq.(\ref{X1120fit}) of unknown form. Mutual interaction
between the surfaces mediated by the bulk may also cause systematic errors.
With $B^t_{110}$ and $\eta_{||,eff}(J_1/J)$ taken from Eq.(\ref{X1120fit})
the shape function
\begin{eqnarray}
\label{S1p20fit}
S^{\alpha}_1(p,L) &=& B^{\alpha}_{\chi_{11}}
{c^{\alpha} - 2\sin(p/2) \over c^{\alpha} + 2\sin(p/2)} \\
&\times& \left[ {2 \Gamma^{\alpha} \sin(p/2) \over \tanh \left(2
\Gamma^{\alpha} \sin(p/2) L \right)} \right]^{\eta_{||,eff}(J_1/J)-1},
\nonumber
\end{eqnarray}
is fitted to $S^t_1$, where the remaining two parameters $c^{\alpha}$ and
$\Gamma^{\alpha}$ are used. As described above the fit is only performed
for $L = 72$ (short dashed line in Fig.\ref{S1p20}). For $L = 24 (\times)$
and $L = 48 (+)$ Eq.(\ref{S1p20fit}) ($\alpha = t$) is shown for $L = 24$
(solid line) and $L = 48$ (long dashed line), where all fit parameters are
kept fixed. The shape function given by Eq.(\ref{S1p20fit}) is remarkably
accurate. However, the observed scaling is completely different from the
asymptotic scaling shown in Fig.\ref{S1p05} and, apart from surface
enhancement corrections, in Fig.\ref{S1p10}. The effective exponent
$\eta_{||,eff}(J_1/J)$ is not related to the surface exponent $\eta_{||}$
given by Eq.(\ref{etapp}), because it is nonuniversal, i.e., it
depends on $J_1/J$. From general considerations it is tempting to pose
the (effective) surface scaling law $\beta_{1,eff}(J_1/J) = \nu [d - 2
+ \eta_{||,eff}(J_1/J)]/2$ (see Eq.(\ref{scalrel})). The direct
determination of $\beta_{1,eff}$ from $m_1(L)$ for $J_1/J \geq 2.0$ is
plagued with considerable uncertainties, because the decay of $m_1(L)$
with $L$ becomes quite slow for $J_1/J \geq 2.0$. Systematic errors
may exceed the formal statistical error of a fit in this case and
therefore the aforementioned effective scaling law cannot be confirmed
unambigiously. Note that the crossover regime to the asymptotic
scaling remains unaccessible within the range of system sizes used
here. Nonetheless, the accuracy of the scaling law for $S_1(p,L)$
for the nonasymptotic regime still lacks theoretical understanding.

For $J_1/J = 1.5$ none of the above descriptions applies to the data
and scaling appears to be violated. This is in accordance with the
findings of Sec. 3, where $m_1(L)$ and $\chi_1(L)$ display sizable deviations
from simple scaling laws. The effects of the crossover to the ordinary
surface universality class are particularly pronounced in the shape
function of the order parameter profile to which we turn in the
following section.

\section{Profiles}
\begin{figure}
\centerline{\epsfig{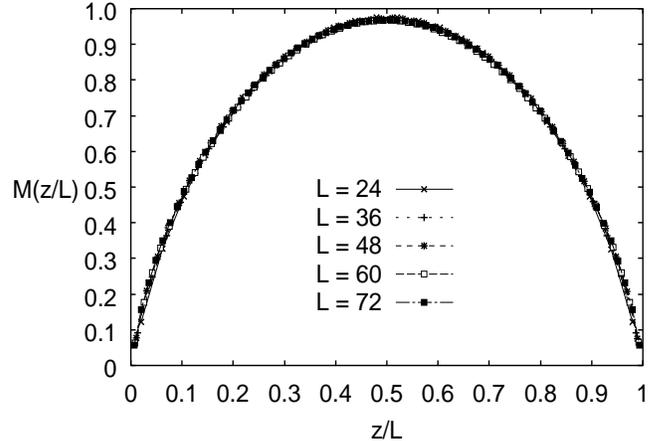}}
\caption{Scaling function $M(z/L)$ of the magnetization profile
(see Eq.(\protect\ref{mscal})) at $T=T_c$ for $J_1/J = 0.5$ and $L = 24$
($\times$, solid line), 36 (+, short dashed line), 48 ($*$, dashed
line), 60 ($\Box$, long dashed line), and 72 (\protect\rule{2mm}{2mm},
dash-dotted line). Statistical errors are much smaller than the symbol
sizes and the various lines are just guides to the eye. Scaling is
obtained in accordance with the ordinary surface universality class.
\label{mz05}}
\end{figure}

The order parameter profile $m(z)$ (see Eq.(\ref{mz})) provides local
information about the order in the system. Furthermore, $\langle m(z)
\rangle$ is by construction very sensitive to the system size at $T=T_c$
and should therefore be a valuable probe for scaling behavior which is
easier to interpret than the structure factor. For $J_1/J \leq
1$ all previous investigations have shown that the system essentially
displays the asymptotic critical behavior of the ordinary surface
universality class. The magnetization profile confirms this
again, so the scaling analysis can be restricted to the case $J_1/J =
0.5$. The scaled magnetization profile $M(z/L)$ is defined by
\begin{equation}
\label{mscal}
M(z/L) \equiv \langle m(z) \rangle / m_b,
\end{equation}
where
\begin{equation}
\label{mb}
m_b \equiv \langle m(L/2) \rangle = B_{m_b} L^{-\beta/\nu}
\left(1 + C_{m_b} L^{-\omega}
\right).
\end{equation}
The factor in parenthesis captures corrections to scaling and the
exponents $\beta$, $\nu$, and $\omega$ are taken from Ref.\cite{GZJ98}.
The coefficient $C_{m_b}$ is determined by a least square fit of Eq.(\ref{mb})
to $m_b$. For numerical convenience $z - 1/2$, $z = 1, \dots, L$ is
chosen as the position coordinate for the profile. The scaling plot
for $J_1/J = 0.5$ is shown in Fig.\ref{mz05}, where $z$ now refers to the
{\em shifted} layer index. The data scale very accurately and the shape of
the profile is in accordance with the expectation for the ordinary
transition. The scaling function $M(z/L)$ can be represented
by the simple fit formula
\begin{equation}
\label{mzfit}
M(\zeta) = B_M \left[ (\zeta + \zeta_0)
(1 - \zeta + \zeta_0) \right]^{(\beta_1 - \beta)/\nu},
\end{equation}
where $\zeta_0 = z_0/L$
and $z_0$ is the extrapolation length, $\beta$ and $\nu$ are taken from
Ref.\cite{GZJ98}, and $\beta_1$ is taken from Eq.(\ref{beta1}). From a
least square fit of Eq.(\ref{mzfit}) to the data for $L = 72$ one
finds $z_0 \simeq -0.26$ in units of the lattice constant. For $J_1/J
= 0.3$ and 1.0 one finds $z_0 \simeq -0.34$ and $z_0 \simeq 0.46$,
respectively. The choice $0.3 \leq J_1/J \leq 1$ therefore yields quite
accurate realizations of Dirichlet boundary conditions, which characterize
the ordinary surface universality class.
\begin{figure}
\centerline{\epsfig{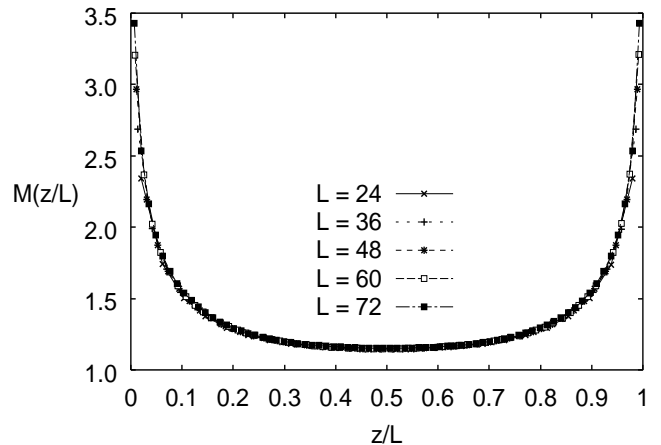}}
\caption{Scaling function $M(z/L)$ of the magnetization profile
(see Eq.(\protect\ref{mscal})) at $T=T_c$ for $J_1/J = 2.0$ and $L = 24$
($\times$, solid line), 36 (+, short dashed line), 48 ($*$, dashed
line), 60 ($\Box$, long dashed line), and 72 (\protect\rule{2mm}{2mm},
dash-dotted line). Statistical errors are much smaller than the symbol
sizes and the various lines are just guides to the eye. The surface
magnetization is strongly enhanced compared to the bulk
magnetization (spurious long - range surface order).
\label{mz20}}
\end{figure}

For $J_1/J = 2.0$ the application of Eqs.(\ref{mscal}) and (\ref{mb})
to the data yields the result shown in Fig.\ref{mz20}. Except at the
surface layers scaling is fulfilled very accurately, however, the
shape of $m(z)$ does not show the expected fixed point form. Instead,
a strong enhancement of the surface magnetization over the
magnetization in the interior is obtained. A behavior like this is
typical for the extraordinary transition which does {\em not} occur
for the Heisenberg model defined by Eq.(\ref{Hamil}) in $d = 3$, i.e.,
the system displays {\em spurious} long-range surface order. If
the lattice size $L$ could be increased further one should therefore
find a crossover to the ordinary behavior displayed in Fig.\ref{mz05}
(see Fig.\ref{m1HI}). Nonetheless, the shape of the scaling function
$M(\zeta)$ is again captured by Eq.(\ref{mzfit}), if the exponent is
used as a third fit parameter. It is tempting to write this exponent
in the form $(\beta_{1,eff}(J_1/J)-\beta)/\nu$, however, the result
for $\beta_{1,eff}(J_1/J)$ obtained this way is not compatible with
the estimate obtained from $m_1(L)$, which may be due to systematic
errors of various kinds discussed above.
\begin{figure}
\centerline{\epsfig{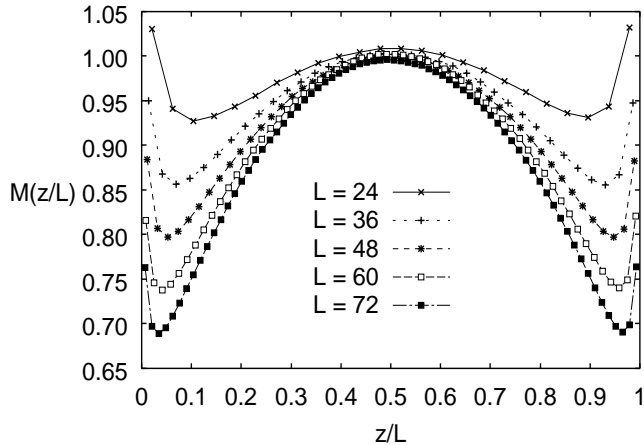}}
\caption{Scaling plot of the magnetization profile $M(z/L)$
(see Eq.(\protect\ref{mscal})) at $T=T_c$ as function of $z/L$ for
$J_1/J = 1.5$ and $L = 24$ ($\times$, solid line), 36 (+, short dashed
line), 48 ($*$, dashed line), 60 ($\Box$, long dashed line), and 72
(\protect\rule{2mm}{2mm}, dash-dotted line). Statistical errors are
much smaller than the symbol sizes and the various lines are just
guides to the eye. Scaling is grossly violated during the crossover
from the state of spurious surface order (see Fig.\protect\ref{mz20})
to the ordinary surface universality class (see
Fig.\protect\ref{mz05}) as $L$ is increased.
\label{mz15}}
\end{figure}

The question how the system actually performs the crossover
is answered in Fig.\ref{mz15}, where $M(z/L)$ is shown for
$J_1/J = 1.5$. For smaller systems the surface magnetization is still
enhanced over $m_b$, but as $L$ is increased, the maximum of the
profile at $z = L/2$ finally exceeds $m_1$ and the profile shape
approaches the fixed point form shown in Fig.\ref{mz05}. The influence
of the surface coupling on $\langle m(z) \rangle$ is confined to the
two outermost lattice layers on either side of the cube, whereas the
curvature of the remainder of the profile already has the ``correct''
sign for all lattice sizes shown in Fig.\ref{mz15}. For $L < 24$ the profile
becomes flatter in the middle and the profile shape approaches the one
displayed in Fig.\ref{mz20}. Note that even at $z = L/2$ the data fail
to collapse according to Eqs.(\ref{mscal}) and (\ref{mb}). Although
the isotropic Heisenberg model does not display long - range surface
order in the thermodynamic limit, on a {\em finite} lattice a strong
enhancement of the surface magnetization $m_1$ over the bulk
magnetization $m_b$ does occur for sufficiently strong surface
couplings as a finite - size effect for a certain range of system sizes.
\begin{figure}
\centerline{\epsfig{figure=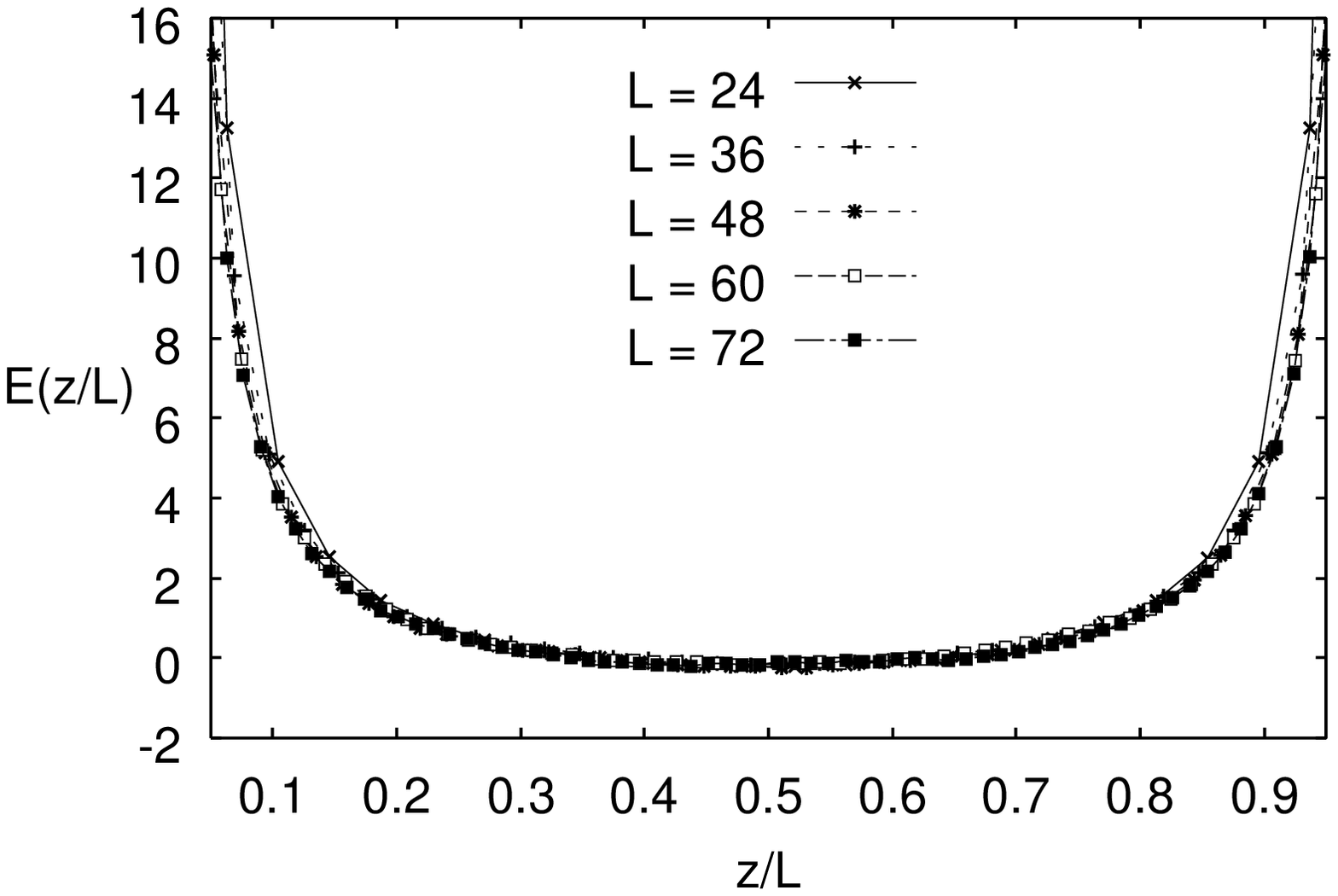,width=6.0cm,height=8.5cm,angle=-90}}
\caption{Scaling function $E(z/L)$ of the energy density profile
(see Eq.(\protect\ref{escal}))
in units of $k_BT_c$ at $T=T_c$ for $J_1/J = 0.5$ and $L = 24$
($\times$, solid line), 36 (+, short dashed line), 48 ($*$, dashed
line), 60 ($\Box$, long dashed line), and 72 (\protect\rule{2mm}{2mm},
dash-dotted line). Statistical errors are smaller than the symbol
sizes and the various lines are just guides to the eye. Scaling is
obtained in accordance with the ordinary surface universality class.
\label{ez05}}
\end{figure}

The scaling properties of the energy density profile are a little more
delicate, because a background energy density must be subtracted from
the profile in order to obtain scaling. One finds the scaling form
\begin{equation}
\label{escal}
E(z/L) \equiv (\langle \varepsilon(z) \rangle - \varepsilon_0) /
L^{-(1-\alpha)/\nu},
\end{equation}
where $E(\zeta)$ is the scaling function and corrections to scaling have
been disregarded. For numerical convenience $z - 1/2$, $z = 1, \dots, L$
is again chosen as the position coordinate for the profile. The scaling
plots for $0.3 \leq J_1/J \leq 1.0$ are well represented by the scaling
plot for $J_1/J = 0.5$ which shown in Fig.\ref{ez05}. As in Figs.\ref{mz05}
- \ref{mz15} $z$ refers to the shifted layer index. The shape of the scaling
function is as expected for the ordinary universality class \cite{KED95}.
For $L > 24$ the data collapse reasonably well which confirms scaling
according to Eq.(\ref{escal}), where $(1-\alpha)/\nu \simeq 1.586$
according to Ref.\cite{GZJ98}. The shape of the scaling function $E(\zeta)$
can be approximated by the fit formula (see also Ref.\cite{KED95})
\begin{equation}
\label{ezfit}
E(\zeta) = B_E \left[ \pi / \sin \left( \pi {\zeta + \zeta_0
\over 1 + 2\zeta_0} \right) \right]^{(1-\alpha)/\nu},
\end{equation}
where $\zeta_0 = z_0/L$ is the scaled extrapolation length of the
profile (see Eq.(\ref{mzfit})). For small arguments $\zeta = z/L$
Eq.(\ref{ezfit}) captures the algebraic increase of $\langle
\varepsilon(z) \rangle$ correctly and the extrapolation length $z_0$
becomes negligibly small. However, away from the surface
Eq.(\ref{ezfit}) captures the shape function of the energy density
profile only in a qualitative sense.
\begin{figure}
\centerline{\epsfig{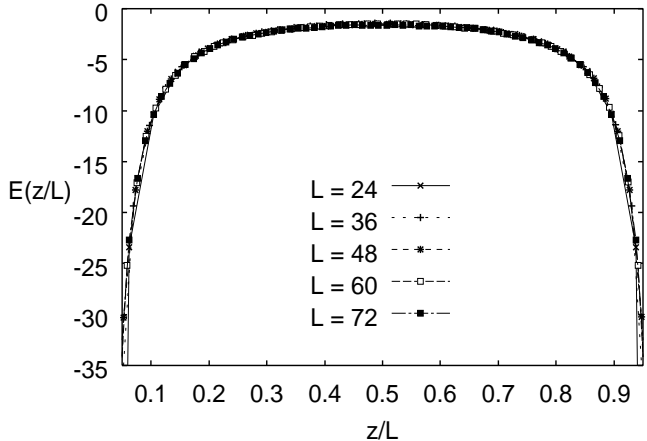}}
\caption{Scaling function $E(z/L)$ of the energy density profile
(see Eq.(\protect\ref{escal}))
in units of $k_BT_c$ at $T=T_c$ for $J_1/J = 2.0$ and $L = 24$
($\times$, solid line), 36 (+, short dashed line), 48 ($*$, dashed
line), 60 ($\Box$, long dashed line), and 72 (\protect\rule{2mm}{2mm},
dash-dotted line). Statistical errors are smaller than the symbol
sizes and the various lines are just guides to the eye. The surface
energy is strongly enhanced compared to the bulk energy (see
Fig\protect\ref{mz20}).
\label{ez20}}
\end{figure}
\begin{figure}
\centerline{\epsfig{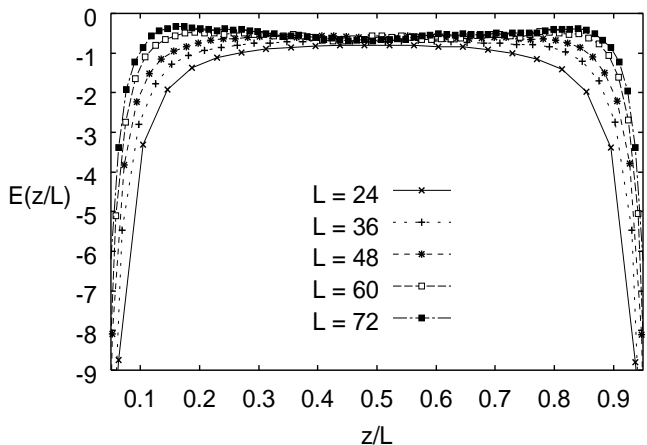}}
\caption{Scaling plot of the energy density profile $E(z/L)$
(see Eq.(\protect\ref{escal})) in units
of $k_BT_c$ at $T=T_c$ as function of $z/L$ for $J_1/J = 1.5$ and $L =
24$ ($\times$, solid line), 36 (+, short dashed line), 48 ($*$, dashed
line), 60 ($\Box$, long dashed line), and 72 (\protect\rule{2mm}{2mm},
dash-dotted line). Statistical errors (not shown)  are slightly larger
than the symbol sizes and the various lines are just guides to the
eye. Scaling is grossly violated during the crossover from the state
of enhanced surface energy (see Fig.\protect\ref{ez20}) to the ordinary
surface universality class (see Fig.\protect\ref{ez05}) as $L$ is
increased.
\label{ez15}}
\end{figure}

For $J_1/J = 2.0$ scaling of the data according to Eq.(\ref{escal})
works even better as shown in Fig.\ref{ez20}. Note that the value
of the reference energy density $\varepsilon_0$ does {\em not}
depend on $J_1/J$. The data for system sizes $L \geq 24$ collapse onto
a single curve, which is represented by Eq.(\ref{ezfit}) with a much
better accuracy than for $J_1/J \leq 1.0$. The extrapolation length
$z_0 \simeq -0.12$ (in units of the lattice constant) is still
very small. The behavior the energy density profile displayed in
Fig.\ref{ez20} is typical for a system at the extrordinary transition
which does not exist for the Heisenberg model in $d = 3$ with nearest
neighbor interactions. According to the above discussion the shape of
the energy density profile given by Fig.\ref{ez20} is governed by the
presence of spurious long - range order in the surface. The crossover
to the aymptotic shape (see Fig.\ref{ez05}) will occur if $L$ is
increased further. For $J_1/J = 2.0$ the crossover regime is out of
reach, but for $J_1/J = 1.5$ this crossover takes place within the
range of accessible system sizes as shown in Fig.\ref{ez15}. As in
Fig.\ref{mz15} scaling is violated. The curvature of the profile near
$z = L/2$ changes sign between $L = 36$ and $L = 60$ and for $L \geq
60$ the profile approaches its asymptotic shape. The nonasymptotic
surface effects are more pronounced here than for the magnetization
profile and penetrate deeper into the system, but the magnitude of the
surface induced enhancement of the energy density at the surface
decays quickly with increasing $L$. The same crossover behavior can be
observed for the critical Ising model slightly below the $SB$ multicritical
point for, e.g., $J_1/J = 1.45$ \cite{MKIsing}.

\section{Summary and outlook}
In the absence of symmetry breaking fields the asymptotic critical
scaling behavior of surfaces of a critical $d = 3$ dimensional Heisenberg
magnet with short - range interactions is always governed by the
ordinary surface universality class. For finite systems, however, a
crucial interplay between the available system size and the value of
the surface - to - bulk coupling ratio $J_1/J$ determines whether or
not the asymptotic surface scaling behavior can actually be observed.
Within the range of system sizes $12 \leq L \leq 72$ for the $L \times
L \times L$ geometry used here critical behavior in the ordinary
surface universality class can be observed for $J_1/J \leq 1.0$. The
scaling exponents found numerically in this case are consistent with
rigorous scaling laws and estimates for theses exponents found
previously by various analytical and numerical methods. The shape of
the surface structure factor $S_1(p,L)$ is captured by very simple
mean - field like expressions, in which two amplitudes are fixed by a
fit to the surface susceptibility for $p = 0$. A width parameter and,
if needed, a surface enhancement parameter then determine the shape of
the momentum dependence. It turns out, that finite - size and lattice
effects are very accurately described by the pseudo scaling argument
$2 \sin(p/2) L$ which replaces the true scaling argument $pL$. These
properties of $S_1(p,L)$ are essential for the data interpretation of
the {\em dynamic} surface structure factor which is the key quantity
for the interpretaion of neutron scattering data on magnetic surfaces
and will therefore be the main focus of ensuing work. The order
parameter and energy density profiles are less relevant for
experiments, but they are easier to interpret and yield valuable
insight into the scaling behavior of the system. Either profile is
found to scale in accordance with the ordinary surface universality
class.

For $J_1/J \geq 2.0$ scaling is still found for all quantities under
investigation, however, the scaling exponents are replaced by
effective ones and their values depend on $J_1/J$. The scaling
relations cannot be verified unambigiously, because some of the
numerical estimates for the effective exponents are presumably
affected by systematic errors of unknown magnitude. Such errors may
ensue due to unknown corrections to the effective scaling
laws or due to an effective interaction between the two surfaces
mediated by the bulk system in between. Nontheless, the effective
scaling properties of the surface structure factor $S_1(p,L)$ provide
valuable information for the analysis of its dynamic counterpart. The
striking scaling properties found here still await theoretical
explanation.

Coupling ratios $J_1/J \geq 2.0$ are too large to access the crossover
regime from the state of enhanced surface magnetization (spurious long
- range surface order) to the asymptotic (ordinary) surface scaling.
For $J_1/J = 1.5$, however, this crossover becomes the dominating
feature in the finite - size behavior of all quantities under
investigation, at least within the range of system sizes used here.
The value $J_1/J = 1.5$ only marks the location of the crossover regime for
the system sizes at hand rather than a sharp transition in the surface
behavior of the Heisenberg model. In the crossover regime scaling is
violated and further theoretical insight is needed for a purposeful
data analysis. Scaling of bulk quantities is also affected by
particularly large correction terms. The crossover process itself is
best visualized in the shape crossover of the magnetization and the
energy density profiles, which occurs as $L$ is increased for $J_1/J =
1.5$ and $T = T_c$.

\acknowledgements
The author gratefully acknowledges financial support of the major part
of this work through the Heisenberg program of the Deutsche
Forschungsgemeinschaft under grant \# Kr 1322/2-1.

\end{document}